\documentclass[twocolumn,amsmath,amssymb,prb]{revtex4-1}
\usepackage{graphicx}
\usepackage{dcolumn}
\usepackage{bm}
\usepackage{color}

\newcommand{\rr}{\textbf{r}}
\newcommand{\xx}{\textbf{r}}

\begin{document}
\title{A Classical Density-Functional Theory for Describing Water Interfaces}

\author{Jessica Hughes}
\affiliation{Department of Physics, Oregon State University,
  Corvallis, OR 97331, USA}
\author{Eric Krebs}
\affiliation{Department of Physics, Oregon State University,
  Corvallis, OR 97331, USA}
\author{David Roundy}
\affiliation{Department of Physics, Oregon State University,
  Corvallis, OR 97331, USA}

\begin{abstract}
We develop a classical density functional for water which combines the
White Bear fundamental-measure theory (FMT)
functional for the hard sphere fluid with
attractive interactions based on the Statistical Associating Fluid
Theory (SAFT-VR).
This functional reproduces the properties of water at both long and
short length scales over a wide range of temperatures, and is
computationally efficient, comparable to the cost of FMT
itself.  We demonstrate our functional by applying it to systems
composed of two hard rods, four hard rods arranged in a square and
hard spheres in water.
\end{abstract}
\maketitle

\section{Introduction}

A large fraction of interesting chemistry---including all of molecular
biology---takes place in aqueous solution.  However, while quantum
chemistry enables us to calculate the ground state energies of
large molecules in vacuum, prediction of the free energy of even the
smallest molecules in the presence of a solvent poses a continuing
challenge due to the complex structure of a liquid and the
computational cost of \emph{ab initio} molecular
dynamics~\cite{car1985, grossman2004}.  The current state-of-the art
in \emph{ab initio} molecular dynamics is limited to a few hundred water
molecules per unit cell~\cite{lewis2011doesnitric}.  On top of this,
standard \emph{ab initio}
methods using classical molecular dynamics without van der Waals
corrections strongly over-structure water, to the point that ice melts
at over 400K\cite{yoo2009phase}!  There has been a flurry of recent
publications implicating van der Waals effects as significant in
reducing this over-structuring\cite{lin2009importance,
  wang2011density, mogelhoj2011ab, jonchiere2011van}.  It has also
been found that the inclusion of nuclear quantum effects can provide
similar improvements \cite{morrone2008nuclear}.  Each of these
corrections imposes an additional computational burden on an approach
that is already feasible for only a very small number of water
molecules. A more efficient approach is needed in order to study
nanoscale and larger solutes.

\subsection{Classical density-functional theory}

Numerous approaches have been developed to approximate the effect of water
as a solvent in atomistic calculations.  Each of these approaches gives an
adequate description of some aspect of interactions with water, but none of
them is adequate for describing \emph{all} these interactions with an
accuracy close to that attained by \emph{ab initio} calculations.  The
theory of Lum, Chandler and Weeks (LCW)~\cite{lum1999hydrophobicity}, for 
instance, can
accurately describe the free energy cost of creating a cavity by placing a
solute in water, but does not lend itself to extensions treating the strong
interaction of water with hydrophilic solutes.  Treatment of water as a
continuum dielectric with a cavity surrounding each solute can give
accurate predictions for the energy of solvation of ions~\cite{latimer1939,
rashin1985, zhan1998, hsu1999, hildebrandt2004, hildebrandt2007}, but
provides no information about the size of this cavity.  In a physically
consistent approach, the size of the cavity will naturally arise from a
balance between the free energy required to create the cavity, the
attraction between the water and the solute, and the steric repulsion which
opens up the cavity in the first place.

One promising approach for an efficient continuum description of water
is that of classical density-functional theory (DFT), which is an
approach for evaluating the free energy and thermally averaged density
of fluids in an arbitrary external potential~\cite{ebner1976density}.
The foundation of classical DFT is the Mermin
theorem\cite{mermin1965thermal}, which extends the Hohenberg-Kohn
theorem\cite{hohenberg1964inhomogeneous} to non-zero temperature,
stating that
\begin{equation}
  A(T) = \underset{n(\rr)}{min}\left\{ F[n(\rr),T] + \int V_\textit{ext}(\rr) n(\rr)
d\rr\right\},
\end{equation}
where $A(T)$ is the Helmholtz free energy of a system in the external
potential $V_\textit{ext}$ at temperature $T$, $n(\rr)$ is the density
of atoms or molecules, and $F[n(\rr),T]$ is a
universal free-energy functional for the fluid, which is independent
of the external potential $V_\textit{ext}$.  Classical DFT is a
natural framework for creating a more flexible theory of
hydrophobicity that can readily describe interaction of water with arbitrary
external potentials---such as potentials describing strong
interactions with solutes or surfaces.

A number of exact properties are easily achieved in the
density-functional framework, such as the contact-value theorem, which
ensures a correct excess chemical potential for small hard solutes.
Much of the research on classical density-functional theory has
focused on the hard-sphere fluid~\cite{curtin1985, rosenfeld1989,
  rosenfeld1993, rosenfeld1997, tarazona1997, tarazona2000}, which has
led to a number of sophisticated functionals, such as the
fundamental-measure theory (FMT) functionals\cite{rosenfeld1989,
  rosenfeld1993, rosenfeld1997, tarazona1997, tarazona2000,
  roth2002whitebear}.  These functionals are entirely expressed as an integral of
local functions of a few convolutions of the density (fundamental
measures) that can be efficiently computed.  We will use the White
Bear version of the FMT functional\cite{roth2002whitebear}.  This
functional reduces to the Carnahan-Starling equation of state in the
homogeneous limit, and it reproduces the exact free energy in the
strongly-confined limit of a small cavity.

A number of classical density functionals have been developed for
water~\cite{ding1987, Yang1992, yang1994density, gloor2002saft,
  gloor2004accurate, gloor2007prediction, Jaqaman2004,
  clark2006developing, chuev2006,
  lischner2010classical, fu2005vapor-liquid-dft,kiselev2006new,
  blas2001examination}, each of which
captures some of the qualitative behavior of water.  However, each of
these functionals also fail to capture some of water's unique
properties.  For instance, the functional of Lischner \emph{et
  al}\cite{lischner2010classical} treats the surface tension
correctly, but can only be used at room temperature, and thus captures
none of the temperature-dependence of water.  A functional by Chuev
and Skolov\cite{chuev2006} uses an ad hoc modification of FMT that can
predict hydrophobic hydration near temperatures of 298~K, but does not produce a
correct equation of state due to their method producting a high value
for pressure. A number of classical density functionals have recently been produced that are based on
Statistical Associating Fluid Theory (SAFT)\cite{ 
  yu2002fmt-dft-inhomogeneous-associating,
  fu2005vapor-liquid-dft,gloor2002saft,muller2001molecular,
  clark2006developing, gloor2007prediction, gloor2004accurate,
  gross2009density, kahl2008modified, blas2001examination}.  These
functionals are based on a perturbative thermodynamic expansion, and
do reproduce the temperature-dependence of water's properties.

\subsection{Statistical associating fluid theory}

Statistical Associating Fluid Theory (SAFT) is a theory describing
complex fluids in which hydrogen bonding plays a significant
role\cite{muller2001molecular}.  SAFT is used to accurately model the
equations of state of both pure fluids and mixtures over a wide range
of temperatures and pressures.  SAFT is
based on Wertheim's first-order thermodynamic perturbation theory
(TPT1)\cite{wertheim1984fluidsI, wertheim1984fluidsII,
  wertheim1986fluidsIII, wertheim1986fluidsIV}, which allows it to
account for strong associative interactions between molecules.

\begin{figure}
\begin{center}
\includegraphics[width=\columnwidth]{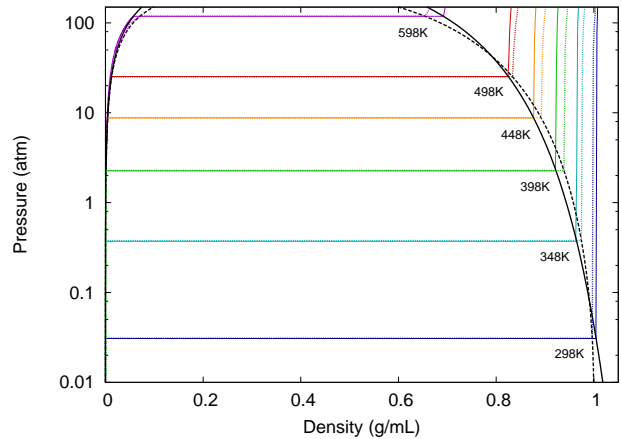}
\end{center}
\caption{The pressure versus density for various temperatures, including
experimental pressure data from NIST\cite{nistwater}. The solid colored lines
indicate the computationally calculated pressure and the dotted
colored lines are NIST data points. The solid and dotted black lines
represent the theoretical and experimental coexistence curves.}
\label{fig:pressure-with-isotherms}
\end{figure}

The SAFT Helmholtz free energy is composed of five terms:
\begin{equation} \label{eq:SAFT-free-energy}
  F = F_\textit{id} + F_\textit{hs} + F_\textit{disp} +
  F_\textit{assoc} + F_\textit{chain},
\end{equation}
where the first three terms---ideal gas, hard-sphere repulsion and
dispersion---encompass the \emph{monomer} contribution
to the free energy, the fourth is the \emph{association} free energy,
describing hydrogen bonds, and the final term is the chain formation
energy for fluids that are chain polymers.  While a
number of formulations of SAFT have been published, we will focus on
SAFT-VR\cite{gil-villegas-1997-SAFT-VR}, which was used by Clark
\emph{et al} to construct an optimal SAFT model for
water\cite{clark2006developing}. All but one of the six empirical
parameters used in the functional introduced in this paper are taken
directly from this Clark \emph{et al} paper.  As an example of the
power of this model, it predicts an enthalpy of vaporization at
100$^\circ$C of $\Delta H_{vap}$= 39.41~kJ/mol, compared with the
experimental value $\Delta H_{vap}$= 40.65~kJ/mol\cite{nistwater},
with an error of only a few percent.  We show a phase diagram for
this optimal SAFT model for water in
Figure~\ref{fig:pressure-with-isotherms}, which demonstrates that its
vapor pressure as a function of temperature is very accurate, while
the liquid density shows larger discrepancies.  The critical point is
very poorly described, which is a common failing of models that are
based on a mean-field expansion.

SAFT has been used to construct classical density functionals,
which are often used to study the surface tension as a function of
temperature\cite{clark2006developing, gloor2004accurate, kahl2008modified,
  gloor2007prediction, blas2001examination, kiselev2006new,
  gloor2002saft,fu2005vapor-liquid-dft, gross2009density,
  yang1994density, Jaqaman2004}.  Such
functionals have qualitatively predicted the dependence of surface
tension on temperature, but they also overestimate
the surface tension by about 50\%, and most SAFT-based functionals are
unsuited for studying systems that have density variations on a
molecular length scale due to the use of a local density
approximation\cite{gloor2002saft,clark2006developing, gloor2007prediction,
gloor2004accurate, gross2009density, kahl2008modified,
blas2001examination, kiselev2006new}.


Functionals constructed using a local density approximation
fail to satisfy the contact-value theorem, and
therefore incorrectly model small hard solutes.
The contact-value theorem states that the pressure any fluid exerts
on a hard wall interface is proportional to the contact density of the
fluid\cite{henderson1979exact}:
\begin{equation}\label{eq:contact}
  p = n_ck_BT,
\end{equation}
where $n_c$ is the contact density, $k_B$ is the Boltzmann constant
and $T$ is the temperature of
the fluid. This leads to the property that the excess chemical
potential of a small hard solute is proportional to the
solvent-excluded volume:
\begin{equation}\label{contactvaluethm}
  F = n k_BT V.
\end{equation}
The contact-value theorem is satisfied by classical density
functionals in which the only purely local term is the ideal gas
contribution to the free energy, and conversely, this theorem is not
satisfied by functionals built using a local density approximation.

\section{Theory and Methods}
We construct a classical density functional for water, which reduces
in the homogeneous limit to the optimal SAFT model for water found by
Clark~\emph{et al}.  The Helmholtz free energy is constructed using the
first four terms from Equation \ref{eq:SAFT-free-energy}: $F_\textit{id}$,
$F_\textit{hs}$, $F_\textit{disp}$ and $F_\textit{assoc}$.  In the
following sections, we will introduce the terms of this functional.

\subsection{Ideal gas functional}
The first term is the ideal gas free energy functional, which is
purely local:
\begin{equation}\label{idealgas}
  F_{id}[n] = k_B T \int n(\xx)\left( \ln{\frac{n(\xx)}{n_Q}} - 1\right) d\xx,
\end{equation}
where n(\xx) is the density of water molecules and $n_Q$ is the
quantum concentration
\begin{equation}\label{quantumconcentration}
 n_Q =\left(\frac{mk_BT}{2\pi\hbar^2}\right)^{3/2}.
\end{equation}
The ideal gas free energy functional on its own satisfies the contact
value theorem and its limiting case of small solutes (Equations
\ref{eq:contact} and \ref{contactvaluethm}). These properties are
retained by our total functional, since all the remaining terms are
purely nonlocal.

\subsection{Hard-sphere repulsion}

We treat the hard-sphere repulsive interactions using the White Bear version of
the Fundamental-Measure Theory~(FMT) functional for the hard-sphere
fluid~\cite{roth2002whitebear}.  FMT functionals are expressed as the integral of
the \emph{fundamental measures} of a fluid, which provide local
measures of quantities such as the filling fraction, density of
spheres touching a given point and mean curvature.  The hard-sphere
excess free energy is written as:
\begin{equation}
F_{hs}[n] = k_B T \int (\Phi_1(\xx) + \Phi_2(\xx) + \Phi_3(\xx)) d\xx \; ,
\end{equation}
with integrands
\begin{align}
\Phi_1 &= -n_0 \ln\left( 1 - n_3\right)\\
\Phi_2 &= \frac{n_1 n_2 - \mathbf{n}_{V1} \cdot\mathbf{n}_{V2}}{1-n_3} \\
\Phi_3 &= (n_2^3 - 3n_2 \mathbf{n}_{V2} \cdot \mathbf{n}_{V2})
  \frac{
    n_3 + (1-n_3)^2\ln(1-n_3)
  }{
    36\pi n_3^2\left( 1 - n_3 \right)^2
  } ,
\end{align}
where the fundamental measure densities are given by:
\begin{align}
  n_3(\xx) &= \int n(\xx') \Theta(\left|\xx - \xx'\right| - R) d\xx' \\
  n_2(\xx) &= \int n(\xx') \delta(\left|\xx - \xx'\right| - R) d\xx'
  \\
  \mathbf{n}_{V2} &= \mathbf{\nabla} n_3 \\
  n_1 &= \frac{n_2}{4\pi R}\\
  \mathbf{n}_{V1} &= \frac{\mathbf{n}_{V2}}{4\pi R}\\
  n_0 &= \frac{n_2}{4\pi R^2}.
\end{align}
The density $n_3$ is the filling fraction and $n_2$ describes the number
of spheres touching a given point. For our functional for water, we use the 
hard-sphere radius of
3.03420~\AA, which was found to be optimal by Clark
\emph{et al}.\cite{clark2006developing}

\newcommand\etadisp{\ensuremath{\eta_\textit{d}}}
\newcommand\epsilondisp{\ensuremath{\epsilon_\textit{d}}}
\newcommand\epsilonassoc{\ensuremath{\epsilon_\textit{a}}}
\newcommand\kappaassoc{\ensuremath{\kappa_\textit{a}}}
\newcommand\lambdadisp{\ensuremath{\lambda_\textit{d}}}
\newcommand\lscale{\ensuremath{s_d}}
\subsection{Dispersion free energy}
The dispersion free energy includes the van der Waals attraction and
any orientation-independent interactions. We use a dispersion term
based on the SAFT-VR approach\cite{gil-villegas-1997-SAFT-VR}, which
has two free parameters (taken from Clark~\emph{et
  al}\cite{clark2006developing}): an interaction energy $\epsilondisp$
and a length scale $\lambdadisp R$.

The SAFT-VR dispersion free energy has the form~\cite{gil-villegas-1997-SAFT-VR}
\begin{align}
  F_\text{disp}[n] &= \int \left(a_1(\xx) + \beta a_2(\xx)\right)n(\xx)d\xx,
\end{align}
where $a_1$ and $a_2$ are the first two terms in a high-temperature
perturbation expansion and $\beta=1/k_BT$.  The first term, $a_1$, is 
the mean-field dispersion interaction. The second term, $a_2$, describes the
effect of fluctuations resulting from compression of the fluid due
to the dispersion interaction itself, and is approximated
using the local compressibility approximation (LCA), which
assumes the energy fluctuation is simply related to the
compressibility of a hard-sphere reference fluid\cite{barker1976liquid}.

The form of $a_1$ and $a_2$ for SAFT-VR is given in
reference~\cite{gil-villegas-1997-SAFT-VR}, expressed in terms
of the filling fraction.  In order to apply this form to an
\emph{inhomogeneous} density distribution, we construct an effective local
filling fraction for dispersion $\etadisp$, given by a Gaussian
convolution of the density:
\begin{align}
  \etadisp(\xx) &= \frac{1}{6\sqrt{\pi} \lambdadisp^3\lscale^3}
  \int n(\xx')\exp\left({-\frac{|\xx-\xx'|^2}{2(2 \lambdadisp
      \lscale R)^2}}\right)d\xx'.
\end{align}
This effective filling fraction is used throughout the dispersion
functional, and represents a filling fraction averaged over the
effective range of the dispersive interaction.  Here we have
introduced an additional empirical parameter $\lscale$ which modifies
the length scale over which the dispersion interaction is correlated.

\subsection{Association free energy}
The final attractive energy term is the association term, which
accounts for hydrogen bonding.  Hydrogen bonds are modeled as four
attractive patches (``association sites'') on the surface of the hard
sphere.  These four sites represent two protons and two electron lone
pairs.  There is an attractive energy $\epsilonassoc$ when
two molecules are oriented such that the proton of one overlaps
with the lone pair of the other.  The volume over which this
interaction occurs is $\kappaassoc$, giving the association
term in the free energy two empirical parameters that are fit to
the experimental equation of state of water (again, taken from
Clark~\emph{et al}\cite{clark2006developing}).

The association functional we use is a modified version of Yu
and Wu\cite{yu2002fmt-dft-inhomogeneous-associating}, which
includes the effects of density inhomogeneities in the contact value
of the correlation function $g^{HS}_\sigma$, but is based on the
SAFT-HS model, rather than the SAFT-VR
model\cite{gil-villegas-1997-SAFT-VR}, which is used in the optimal
SAFT parametrization for water of Clark \emph{et
  al}\cite{clark2006developing}.  Adapting Yu and Wu's association
free energy to SAFT-VR simply involves the addition of a correction
term in the correlation function (see Equation~\ref{eq:gSW}).

The association functional we use is constructed by using the density
$n_0(\xx)$, which is the density of hard spheres touching a given
point, in the standard SAFT-VR association
energy\cite{gil-villegas-1997-SAFT-VR}.
The association free energy for our four-site model has the form
\begin{align}
  F_\text{assoc}[n] &= 4 k_BT \int n_0(\xx) \zeta(\xx)
  \left(\ln X(\xx) - \frac{X(\xx)}{2} + \frac12\right) d\xx,
\end{align}
where the factor of $4$ comes from the four association sites per
molecule, the functional $X$ is the fraction of association sites
\emph{not} hydrogen-bonded, and $\zeta(\xx)$ is a dimensionless
measure of the density inhomogeneity.
\begin{align}
  \zeta(\xx) &= 1 - \frac{\mathbf{n}_{2V}\cdot\mathbf{n}_{2V}}{n_2^2}.
\end{align}
The fraction $X$ is determined by the quadratic equation
\begin{align}
  X(\xx) &= \frac{\sqrt{1 + 8n_0(\xx)\zeta(\xx)
      \Delta(\xx)} - 1}
  {4 n_0(\xx)\zeta(\xx)
    \Delta(\xx)},
\end{align}
where the functional $\Delta$ is a measure of hydrogen-bonding
probability, given by
\begin{align}
  \Delta(\xx) &= \kappaassoc g^\textit{SW}_\sigma(\xx)
  \left(e^{-\beta\epsilonassoc} - 1\right) \\
  g^\textit{SW}_\sigma(\xx) &= g^\textit{HS}_\sigma(\xx) +
  \frac{1}{4}\beta\left(\frac{\partial a_1}{\partial \etadisp(\xx)} -
  \frac{\lambdadisp}{3 \etadisp}\frac{\partial a_1}{\partial \lambdadisp}\right)\label{eq:gSW},
\end{align}
where $g^\textit{SW}_\sigma$ is the correlation function evaluated at
contact for a hard-sphere fluid with a square-well dispersion
potential, and $a_1$ and $a_2$ are the two terms in the dispersion
free energy.  The correlation function $g^\textit{SW}_\sigma$ is
written as a perturbative correction to the hard-sphere fluid
correlation function $g^\textit{HS}_\sigma$, for which we use the
functional of Yu and Wu\cite{yu2002fmt-dft-inhomogeneous-associating}:
\begin{align}
  g_\sigma^{HS} &= \frac{1}{1-n_3}
  +\frac{R}{2}\frac{\zeta n_2}{(1-n_3)^2}
  + \frac{R^2}{18}\frac{\zeta n_2^2}{(1-n_3)^3}.
\end{align}

\subsection{Determining the empirical parameters}\label{sec:empirical}

\begin{figure}
\begin{center}
\includegraphics[width=\columnwidth]{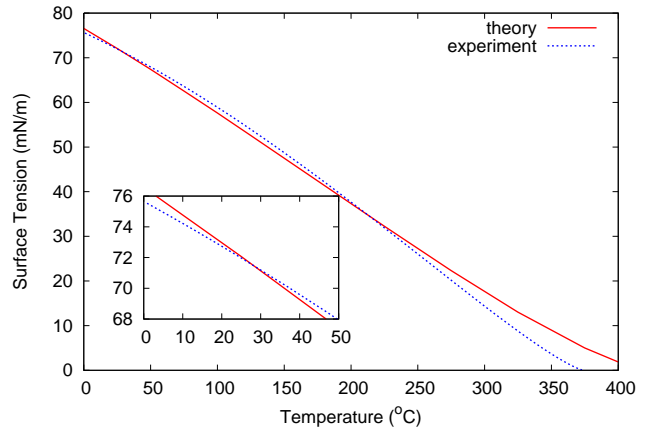}
\end{center}
\caption{Comparison of Surface tension versus temperature for theoretical and
  experimental data. The experimental data is taken from NIST.\cite{nistwater}
  The length-scaling parameter $\lscale$ is fit so that the theoretical surface 
  tension will match the experimental surface tension near room temperature.}
\label{fig:surface-tension}
\end{figure}

The majority of the empirical parameters used in our functional are
taken from the paper of Clark \emph{et al} on developing an optimal
SAFT model for water\cite{clark2006developing}.  This SAFT model
contains five empirical parameters: the hard-sphere radius, an energy
and length scale for the dispersion interaction, and an energy and
length scale for the association interaction.  In addition to the five
empirical parameters of Clark \emph{et al}, we add a single additional dimensionless
parameter $\lscale$---with a fitted value of 0.353---which determines
the length scale over which the density is averaged when computing the
dispersion free energy and its derivative.  We determine this final
parameter by fitting the to the experimental surface tension with the
result shown in Figure~\ref{fig:surface-tension}.  Because the SAFT
model of Clark \emph{et al} overestimates the critical
temperature---which is a common feature of SAFT-based functionals that
do not explicitly treat the critical point---we cannot reasonably
describe the surface tension at all temperatures, and choose to fit
the surface tension at and around room temperature.

\section{Results and discussion}

\subsection{One hydrophobic rod}

\begin{figure}
\begin{center}
\includegraphics[width=\columnwidth]{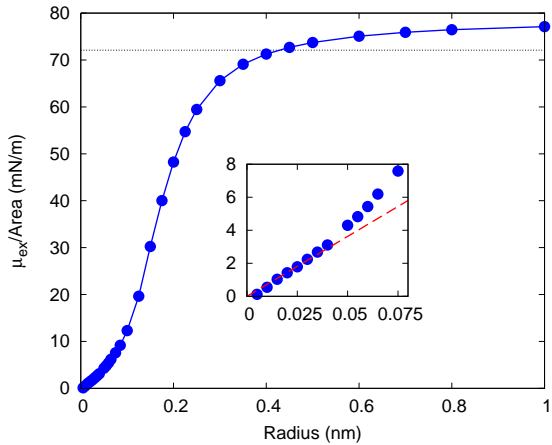}
\end{center}
\caption{ Excess chemical potential per area versus radius for a single
hydrophobic rod immersed in water. This should have an asymptote equal
to the surface tension at room temperature, and it agrees well with
the surface tension in Figure~\ref{fig:surface-tension}. The inset for
rods with a very small radius shows the linear relationship expected based on
Equation~\ref{contactvaluethm}.}
\label{fig:energy-vs-diameter}
\end{figure}

We begin by studying a single hydrophobic rod immersed in water.
In Figure~\ref{fig:energy-vs-diameter} we show the excess chemical
potential at room temperature, scaled by the solvent-accessible
surface area of the hard rod, plotted as a function of hard-rod
radius.  We define the hard-rods radius as the radius from which water
is excluded.
For rods with radius larger than 0.5~nm or so, this reaches a maximum value of
75~mN/m, which is slightly higher than the bulk surface.  In the limit
of very large rods, this value will decrease and approach the bulk
surface value.  As seen in the inset of
Fig.~\ref{fig:energy-vs-diameter}, for rods with very small radius (less than
about 0.5~\AA) the excess chemical potential is proportional to volume,
as required by the contact-value theorem (see Equation~\ref{contactvaluethm}).

\begin{figure}
\begin{center}
\includegraphics[width=\columnwidth]{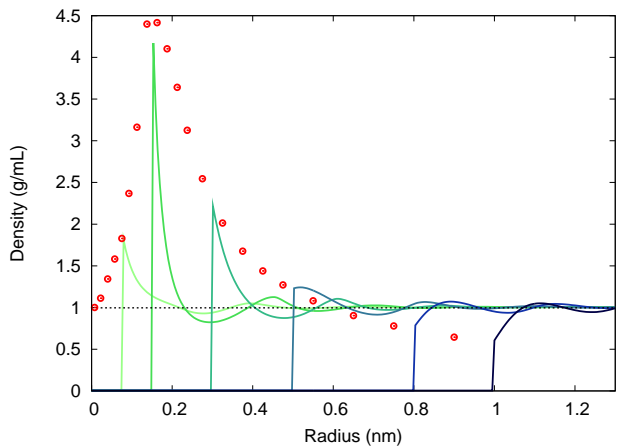}
\end{center}
\caption{ Density profiles for single rods of different radii. The dotted line
represents the saturated liquid density and the points represent the
expected contact density derived from the contact value theorem and
calculated free energy data.}
\label{fig:density-single-rod}
\end{figure}

We show in Figure~\ref{fig:density-single-rod} density profiles for
different radii rods, as well as the prediction for the contact value
of the density as a function of rod radius, as computed from the free
energies plotted in Figure~\ref{fig:energy-vs-diameter}.  The
agreement between these curves confirms that our functional satisfies
the contact-value theorem and that our minimization is well converged.
As expected, as the radius of the rods becomes zero the
contact density approaches the bulk density, and as the radius becomes
large, the contact density will approach the vapor density.

\subsection{Hydrophobic interaction of two rods}

\begin{figure}
\begin{center}
\includegraphics[width=\columnwidth]{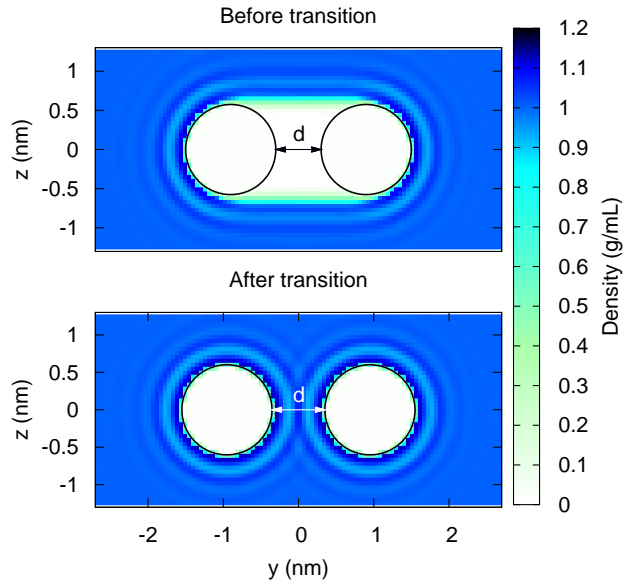}
\end{center}
\caption{ Density profiles illustrating the transition from vapor 
to liquid water between the rods. The radius is 0.6~nm, the top figure is
at a separation of 0.6~nm and the
bottom is 0.7~nm. Figure~\ref{fig:rods-energy-vs-distance} shows
the energy for these and other separations.}
\label{fig:density-rods}
\end{figure}

We now look at the more interesting problem of two parallel hard rods
in water, separated by a distance $d$, as shown in
Figure~\ref{fig:density-rods}.  At small separations there is only vapor
between the rods, but as the rods are pulled apart, the vapor region
expands until a critical separation is reached at which point liquid
water fills the region between the rods.
Figure~\ref{fig:density-rods} shows density profiles before and after
this transition for rods of radius 0.6~nm. This critical separation
for the transition to liquid depends on the radii of the rods, and is
about 0.65~nm for the rods shown in Figure~\ref{fig:density-rods}.
The critical separation will be different for a system where there is
attraction between the
rods and water.  At small separations, the shape of the water around the two
rods makes them appear as one solid ``stadium''-shaped object
(a rectangle with semi-circles on both ends).


To understand this critical separation, we consider the free energy in
the macroscopic limit, which is given by
\begin{equation}
F = \gamma A + pV.
\end{equation}
The first term describes the surface energy and the second term is the
work needed to create a cavity of volume $V$. Since the pressure term
scales with volume, it can be neglected relative to the surface term
provided the length scale is small compared with $\gamma / p$, which
is around $20~\mu m$, and is much larger than any of the systems we
study. For micron-scale rods, the water on the sides of the `stadium'
configuration will bow inward between the rods and the density will
reduce to vapor near the center point where the rods are closest to
each other.

Starting from the surface energy term, we can calculate the 
free energy per length, which is equal to the circumference multiplied 
by the surface tension. The force per length is the derivative of this
with respect to the separation. 
The circumference of the stadium-shape is
\begin{equation}
C_{s} = 2\pi r +4r+2d
\end{equation}
and so the force per length is equal to twice the surface tension.

\begin{figure}
\begin{center}
\includegraphics[width=\columnwidth]{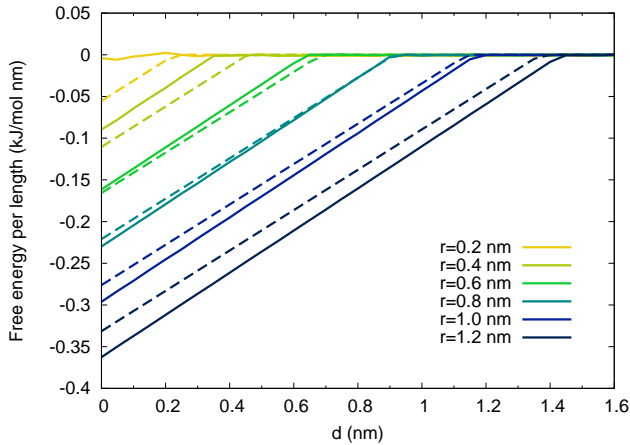}
\end{center}
\caption{ Free energy of interaction (also known as the potential of mean force)
versus separation for two hydrophobic rods ranging in radius from
0.2~nm to 1.2~nm.
All were arbitrarily offset to zero at large separations for ease of comparison. The
transition corresponds to the phase change from
vapor to liquid between the rods as pictured in the density profiles in 
Figure~\ref{fig:density-rods}. }
\label{fig:rods-energy-vs-distance}
\end{figure}

We plot in Figure~\ref{fig:rods-energy-vs-distance} the computed free
energy of interaction
per unit length from our classical density functional (solid lines),
as a function of the separation $d$, along with the free energy
predicted by our simple macroscopic model (dashed lines).  The models
agree very well on the force between the two rods at close
separations, and have reasonable agreement as to the critical
separation for rods greater than 0.5~nm in radius.

Walther \emph{et al}\cite{walther2004hydrophobic} studied the
interactions between two carbon nanotubes, which are geometrically
similar to our hydrophobic rods, using molecular dynamics with the SPC
model for water, and a Lennard-Jones potential for the interaction of
carbon with water, for nanotubes of diameter 1.25~nm and separations
ranging from about 0.3~nm to 1.5~nm.  The SPC model underestimates the
surface tension of water by about 24\%\cite{vega2007surface}, so we
cannot expect this work to provide quantitative agreement with real
water.  Walther \emph{et al} observe a drying transition between the
two nanotubes, which occurs at a smaller radius than our results
suggest.  However, when Walther \emph{et al} disable the attractive
interaction between nanotube and water, the drying effect occurs at
much longer range, in agreement with our results.

\subsection{Hydrophobic interactions of four rods}

\begin{figure}
\begin{center}
\includegraphics[width=\columnwidth]{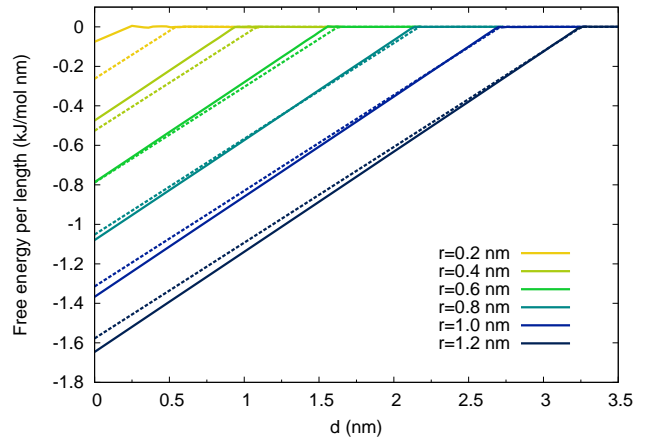}
\end{center}
\caption{ Free energy of interaction
versus separation for four hydrophobic rods ranging in radius from
0.2~nm to 1.2~nm.
All were arbitrarily offset to zero at large separations. The
transition corresponds to the phase change from
vapor to liquid between the rods as pictured in the density profiles in 
Figure~\ref{fig:density-four-rods}. }
\label{fig:four-rods-energy-vs-distance}
\end{figure}

\begin{figure}
\begin{center}
\includegraphics[width=\columnwidth]{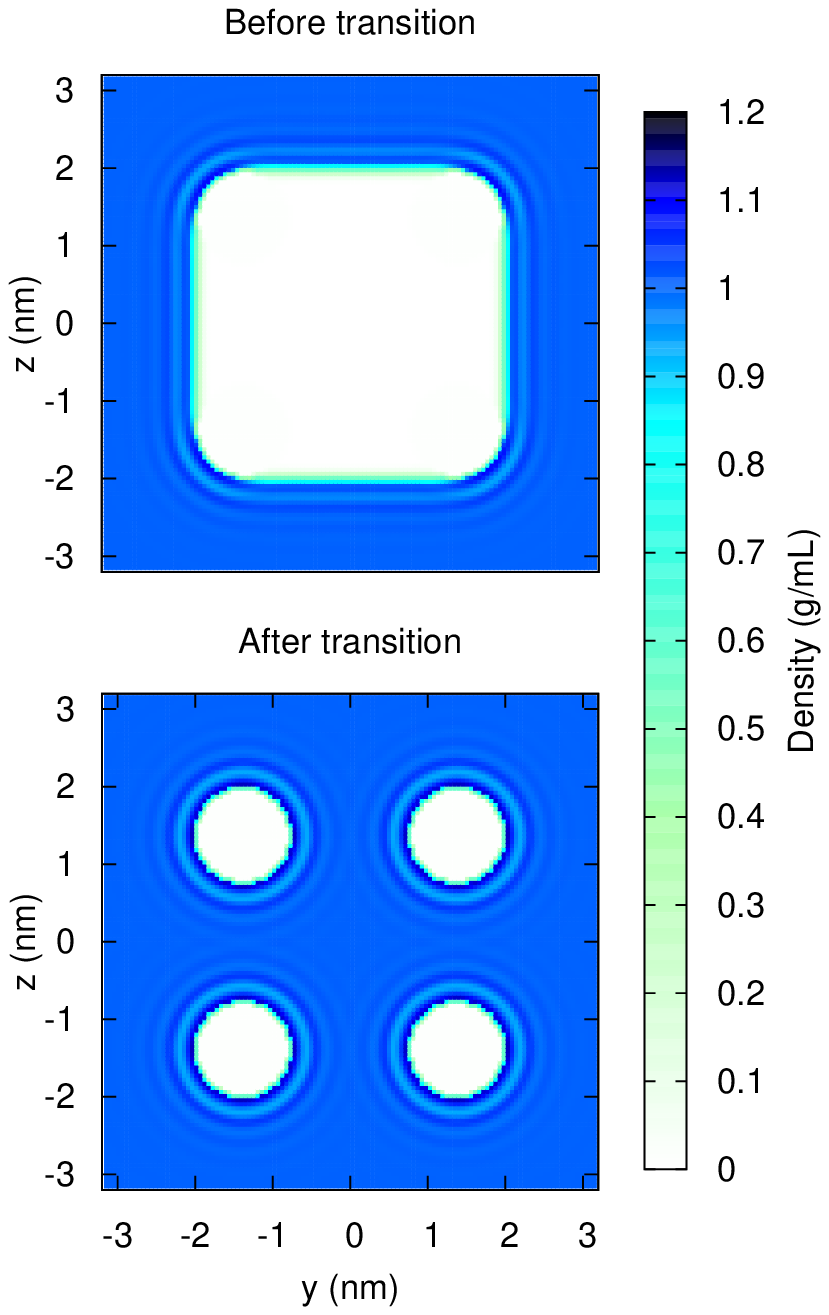}
\end{center}
\caption{ Density profiles illustrating the transition from vapor 
to liquid water between four rods. The radius is 0.6~nm, the top figure is
at a separation of 1.53~nm and the
bottom is 1.56~nm. Figure~\ref{fig:four-rods-energy-vs-distance} shows
the energy for these and other separations.}
\label{fig:density-four-rods}
\end{figure}

We go on to study four parallel hard rods, as examined by Lum,
Chandler and Weeks in their classic paper on hydrophobicity at small
and large length scales~\cite{lum1999hydrophobicity}.  As in the case
of two rods---and as predicted by Lum~\emph{et al}---we observe a
drying transition, as seen the density plot shown in
Figure~\ref{fig:density-four-rods}.  In
Figure~\ref{fig:four-rods-energy-vs-distance}, we plot the free energy
of interaction together with the macroscopic approximation, and find
good agreement for rods larger than 0.5~nm in radius.  This free
energy plot is qualitatively similar to that predicted by the LCW
theory~\cite{lum1999hydrophobicity}, with the difference that we find
no significant barrier to the association of four rods.

\subsection{Hydration energy of hard-sphere solutes}

\begin{figure}
\begin{center}
\includegraphics[width=\columnwidth]{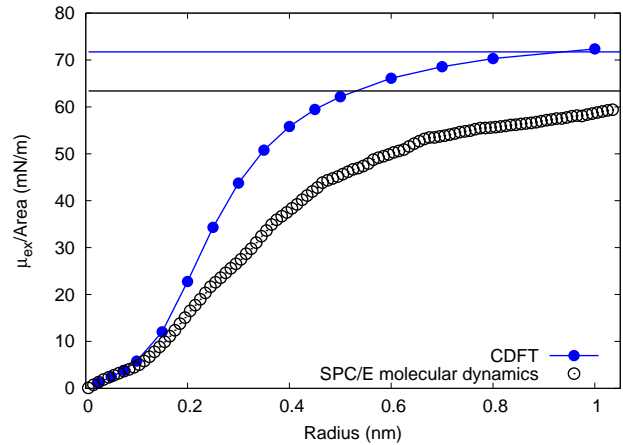}
\end{center}
\caption{ Excess chemical potential per area versus radius for
  a single hydrophobic sphere
  immersed in water. This should have an asymptote equal to the
  surface tension at room temperature, and it agrees well with the
  surface tension in Figure~\ref{fig:surface-tension}. Results from a
  simulation of SPC/E water~\cite{huang2001shs} are shown as circles.
  The horizontal lines show the experimental and SPC/E bulk surface
  tension for water at standard atmospheric temperature and
  pressure. }
\label{fig:sphere-energy-vs-diameter}
\end{figure}

A common model of hydrophobic solutes is the hard-sphere solute, which
is the simplest possible solute, and serves as a test case for
understanding of hydrophobic solutes in
water\cite{sedlmeier2011entropy}.  As in the single rod, we begin by
examining the ratio of the excess chemical potential of the cavity
system to the solvent-accessible surface area
(Figure~\ref{fig:sphere-energy-vs-diameter}).  This
effective surface tension surpasses the bulk surface tension at a
radius of almost 1~nm, and at large radius will drop to the bulk value.  As
with the single rod, we see the analytically correct behavior in the
limit of small solutes.  For comparison, we plot the free energy
calculated using a molecular dynamics simulation of SPC/E
water\cite{huang2001shs}.  The agreement is quite good, apart from the
issue that the SPC/E model for water significantly underestimates the
surface bulk tension of water at room
temperature\cite{vega2007surface}.

\begin{figure}
\begin{center}
\includegraphics[width=\columnwidth]{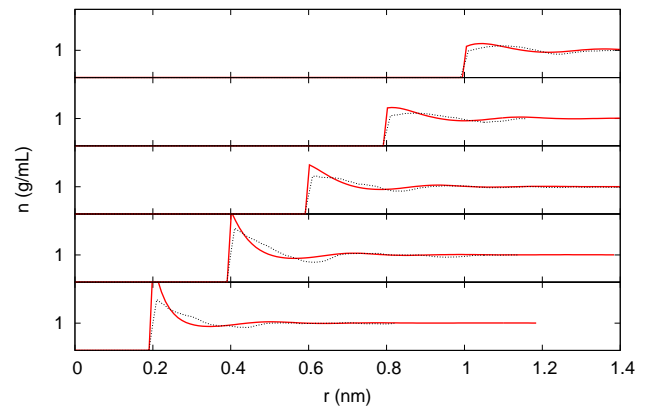}
\end{center}
\caption{ Density profiles around hard-sphere solutes of different radii. Predictions
  from our classical density-functional theory are in solid red, while
  the dotted line shows the result of a molecular dynamics simulation
  of SPC/E water~\cite{huang2001shs}.  }
\label{fig:density-sphere}
\end{figure}

Figure~\ref{fig:density-sphere} shows the density profile for several
hard sphere radii, plotted together with the results of the same
SPC/E molecular dynamics simulation shown in
Figure~\ref{fig:sphere-energy-vs-diameter}\cite{huang2001shs}.  The
agreement with simulation is quite reasonable.  The largest
disagreement involves the density at contact, which according to the
contact value theorem cannot agree, since the free energies do not
agree.

\section{Conclusion}

We have developed a classical density functional for water that
combines SAFT with the fundamental-measure theory for hard spheres,
using one additional empirical parameter beyond those in the SAFT
equation of state, which is used to match the experimental surface
tension.  This functional does \emph{not} make a local density
approximation, and therefore correctly models water at both small and
large length scales.  In addition, like all FMT functionals, this
functional is expressed entirely in terms of convolutions of the
density, which makes it efficient to compute and minimize.

We apply this functional to the case of hard hydrophobic rods and
spheres in water.  For systems of two or four hydrophobic rods
surrounded by water, we see a transition from a vapor-filled state a
liquid-filled state.  A simple model treatment for the critical
separation for this transition works well for rods with diameters
larger than 1~nm.  In the case of spherical solutes, we find good
agreement with SPC/E simulations.

\bibliography{paper}

\end{document}